\documentclass[amsmath,amssymb,floatfix,prb,epsf,superscriptaddress,twocolumn]{revtex4}
\usepackage{amsmath,amssymb,natbib,bm,graphicx,url,epsfig,wasysym}
\usepackage[ansinew]{inputenc}
\usepackage{bm}
\usepackage{color}
\usepackage{appendix}
\newcommand{\be}{\begin{equation}}
\newcommand{\ee}{\end{equation}}
\newcommand{\bes}{\begin{equation}\begin{split}}
\newcommand{\ees}{\end{split}\end{equation}}
\newcommand{\bea}{\begin{eqnarray}}
\newcommand{\eea}{\end{eqnarray}}
\newcommand{\nn}{\nonumber}

\DeclareMathOperator{\curl}{curl}

\def\beq{\begin{equation}}
\def\eeq{\end{equation}}
\def\bea{\begin{eqnarray}}
\def\eea{\end{eqnarray}}
\def\ena{\end{eqnarray}}

\begin{document}

%\maketitle
%\title{$\mathbb{Z}_2$ Breaking Dirac spin-liquid  in a spin-$\frac{1}{2}$ $J_1-J_2$ triangular antiferromagnet}
%\title{Nature of the Spin Liquid  in  Frustrated Quantum  XXZ Magnets}
\title{Helical spin liquid in a triangular XXZ magnet from Chern-Simons theory}

\author{Tigran Sedrakyan}

\affiliation{Precision Many-Body Physics Initiative, Physics Department, University of Massachusetts, Amherst, Massachusetts 01003, USA}
\affiliation{Max-Planck-Institut f{\"u}r Physik komplexer Systeme, N{\"o}thnitzer Straße 38, D-01187 Dresden, Germany}

\author{Roderich Moessner}
\affiliation{Max-Planck-Institut f{\"u}r Physik komplexer Systeme, N{\"o}thnitzer Straße 38, D-01187 Dresden, Germany}

\author{Alex Kamenev}
\affiliation{William I. Fine Theoretical Physics Institute,  School of Physics and Astronomy, University of Minnesota, Minneapolis, Minnesota 55455, USA}

\begin{abstract} 
We propose a finite-temperature phase diagram  for the 2D spin-$1/2$ $J_1-J_2$ $XXZ$ antiferromagnet on the  triangular lattice. Our analysis, based on 
a composite fermion representation, 
yields several phases. This includes  
a zero-temperature helical spin liquid with 
$N=6$ {\it anisotropic} Dirac cones, and 
with nonzero vector chirality implying a broken $\mathbb{Z}_2$ symmetry. 
It is terminated at $T=0$  by a continuous quantum phase transition to 
$120^\circ$ ordered state around $J_2/J_1\approx0.089$ in the XX limit; these phases share a double degeneracy, which persists to finite $T$ above the helical spin liquid. By contrast, 
at $J_2/J_1 \simeq 0.116$, the transition into a stripe phase appears as first order. 
We further discuss experimental and numerical  consequences of the helical order and the anisotropic 
nature of the Dirac dispersion. 
\end{abstract}

\date{\today}

\maketitle

\section{Introduction} 
Two-dimensional $s=1/2$ magnets with  frustrated interactions attract a great deal of interest because of their 
potential to host unconventional states of quantum matter such as  spin liquids (SL) \cite{FA,balents,kitaev,savary,norman,XGW,KM,ng,senthil}
%\cite{balents,kitaev,savary,norman,khveshch,XGV,hermele,kivelson,berg,shannon1,lu,pereira}. 
Quantum SL  are long range entangled states that give rise to emergent gauge fields and represent deconfined phases,  where the quasi-particles exhibit fractional quantum numbers. They do not break rotational symmetry, thus excluding  orientational long range ordering. Traditionally, the triangular lattice has been regarded as a promising ground for realization of a SL\cite{Andreson-1973,ss,Moessner1,moess,AV,wietek}.
In this setting, the frustrated spin-$1/2$ $J_1-J_2$ $XXZ$ antiferromagnet on a triangular lattice is one such 
candidates for a SL ground state in a parameter window around $J_2/J_1 \sim 0.1$. The Hamiltonian of the model is given by 
\begin{eqnarray}
\label{H-XXZ}
&&\hat {\mathcal H}=\hat {\mathcal H}_1+\hat {\mathcal H}_2, {\text{where}}\\
&&\hat {\mathcal H}_l  =  \frac{J_l}{2} \sum_{{\bf r},\nu} \big[\hat{S}^{+}_{\bf r}{\hat S}^{-}_{{\bf r}+{\bm \mu}_\nu^{l}}+
\hat{S}^{-}_{\bf r}\hat{S}^{+}_{{\bf r}+{\bm \mu}_\nu^{l}}+2 g \hat{S}^{z}_{\bf r}\hat{S}^{z}_{{\bf r}+{\bm \mu}_\nu^{l}}\big], \nn
\end{eqnarray}
where $l=1,2$ and  a parameter $g$ measures the anisotropy of the interactions.  
Vectors ${\bm \mu}_\nu^1={\bm e}_\nu$  and ${\bm \mu}_\nu^2={\bm a}_\nu, \;  \nu=1,2,3$, point to  nearest neighbor (NN) and  next nearest neighbor (NNN) sites on the triangular lattice, respectively. The spin-orbit coupled version of the model is believed to be related to the triangular lattice antiferromagnet YbMgGaO$_4$
\cite{shen,yli,paddison,chen17,chen18,starykh}.

\begin{figure}[t]
	\center
	%\scalebox{0.5}{\rotatebox{0}{\includegraphics{Mbox.eps}}}
	\centerline{\includegraphics[width=75mm,angle=0,clip]{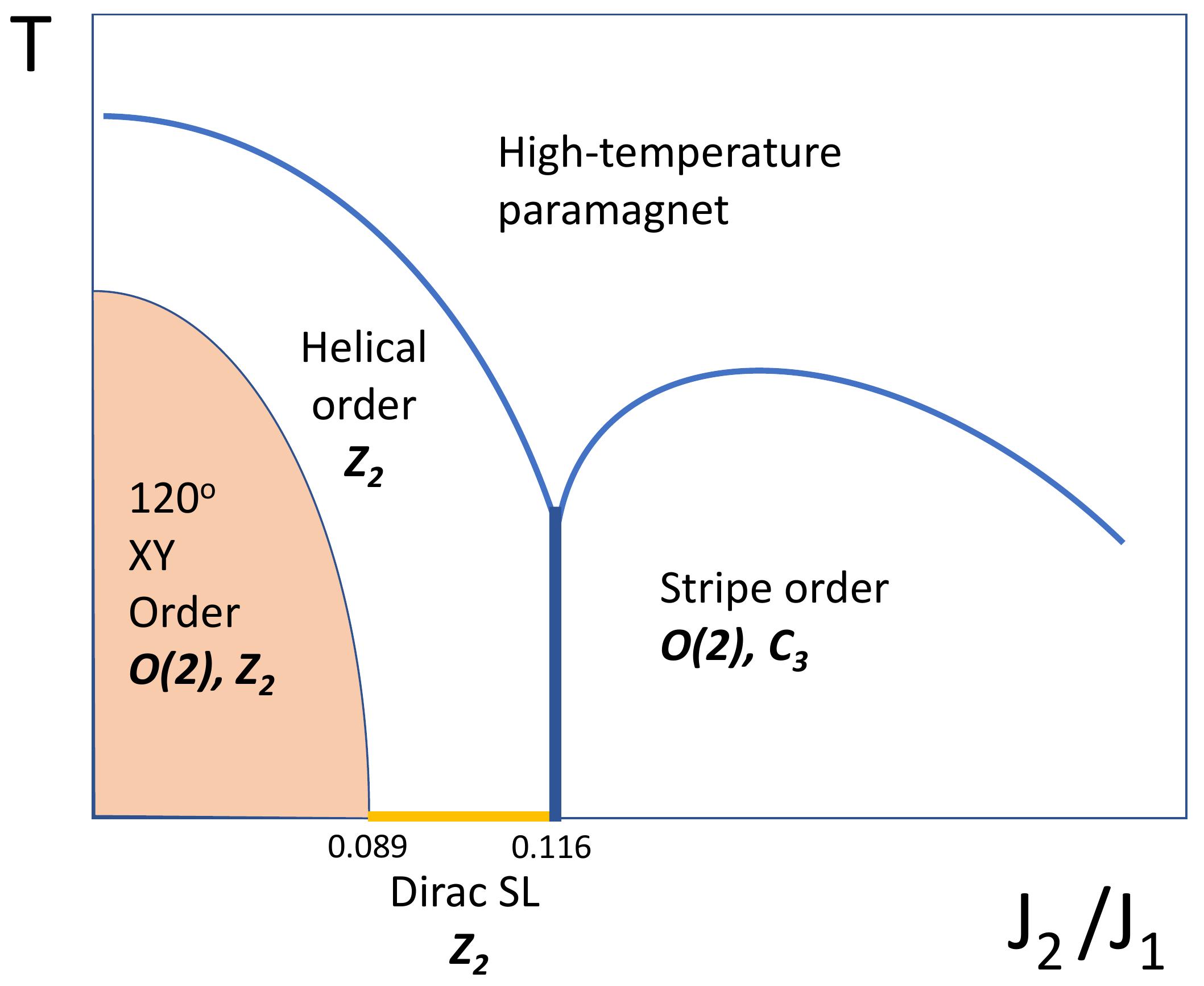}}
	\caption{(Color online) A schematic finite-temperature phase diagram of the $J_1-J_2$ XXZ model on a triangular lattice from our composite Fermion analysis.  Broken symmetries are indicated in bold font. 
The thick  (blue) line represents the first-order transition from helical into stripe ordered phases. Thick horizontal (yellow) line represents the $T=0$ helical Dirac SL. Numerical values for $J_2/J_1$ correspond to the XX case, $g=0$.}
	\label{T-phase}
\end{figure}

The Heisenberg model, $g=1$, has been studied numerically using variational Monte-Carlo~\cite{vmc1,becca,Balents-2018,becca-prx} and  the density matrix renormalization group (DMRG)~\cite{White-2015,Sheng-2015,crit,xxz}. 
The DMRG study of Refs.~[\onlinecite{crit,xxz}] and 
the variational Monte-Carlo study of Ref.~[\onlinecite{Balents-2018}]  explore the phase diagram of the model Eq.~(\ref{H-XXZ})  ranging from the $XX$ limit, $g=0$, all the way to the Heisenberg limit with $g=1$.  The nature of  the SL in Ref.~[\onlinecite{Balents-2018}] was identified with the $U(1)$ Dirac gauge theory which emerges 
in the Heisenberg limit in an approximate parameter interval 
$0.08\lesssim t \lesssim 0.2 $, $t=J_2/J_1$\cite{becca,becca-prx}. The interval where the SL is realized appears to be narrower in the $XXZ$ model\cite{crit,Balents-2018,xxz}. 

In this work we carry out an analysis based on a composite Fermion representation. 
An advantage of the fermion representation is that it can be used  to effectively describe both the ordered phases as CS superconductors\cite{CSS}, 
and spin-liquids, where the fermions can be "deconfined." 
There are two main differences between our work and previous approaches to establish spin-liquids: in our framework, we start by focusing on the ordered states of the spin-1/2 XXZ magnet via treating it as superconducting states of spinless Chern-Simons fermions. We then study the stability region of the ordered state, assuming the spin-liquid emerges when the ordering breaks down. 
Apart from developing the general approach based on breaking of the Chern-Simons superconductivity for detecting instability of the spin-order, we apply the method to the specific model under consideration. The purpose of this application is to propose a new state -- the helical spin liquid.  In part, our work is a study that suggests the existence of new types of phases with peculiar properties, which also indicates where they may be realized. Our approach proposes a novel scenario of unconventional (deconfined) phase transitions and novel quantum phases that can be realized in frustrated magnets.

The performed analysis leads us to propose that the  rotational $O(2)$ symmetry is restored, and a kind of Dirac SL  is stabilized in the XXZ model in a narrow interval of parameter $t=J_2/J_1$, eg. $0.089\lesssim t \lesssim 0.116$ in XX limit. 
The nature of this SL appears to be different from the  Heisenberg limit in  a crucial way: in the XXZ model it exhibits spontaneous breaking of the $\mathbb{Z}_2$ symmetry, inherent to the  
$120^\circ$ antiferromagnetic ordering of the XX model at $J_2=0$. We thus predict a SL with long range $\mathbb{Z}_2$ order and the {\em vector chirality} playing the role of the order parameter which distinguishes between two degenerate SL ground states. The other central finding is that such a SL is described in terms of $N=6$ copies of Dirac fermions. Each Dirac cone is predicted to exhibit a uniaxial anisotropy, although the complete spectrum preserves the $C_3$ symmetry of the lattice. A finite vector chirality and the anisotropy of individual Dirac cones are the main properties of the proposed SL, 
dubbed here {\it Helical} SL, which may be detected using DMRG, tensor network, or variational Monte-Carlo approaches. They may be also observed in spin resolved neutron scattering.

Furthermore, our theory predicts a rich finite-temperature phase diagram, Fig.~\ref{T-phase}. 
As in the classical XX model \cite{Lee,Shiba,korshunov},  a BKT transition takes place first into a helical phase with restored $O(2)$, but broken $\mathbb{Z}_2$ symmetries. 
At yet higher temperature, there is an Ising-like transition to a disordered paramagnet. Finally, we argue that the $T=0$ transition from $120^\circ$ ordered state to the helical SL is continuous, while the one into the stripe phase is first order.

We treat these transitions by first developing a theory of  the $120^\circ$ state via Chern-Simons (CS) superconductivity \cite{CSS}, 
considering the stability of the superconducting solution upon increasing $t$. The superconducting order breaks down at $t \sim 0.089$ for $g=0$, signaling an emergence of the Dirac SL state with broken $\mathbb{Z}_2$ symmetry.  Next, we identify a CS superconductor describing the collinear stripe phase, energetically favorable beyond $t\sim 0.116$. 

\begin{figure}[t]
	\center
	%\scalebox{0.5}{\rotatebox{0}{\includegraphics{Mbox.eps}}}
	\centerline{\includegraphics[width=85mm,angle=0,clip]{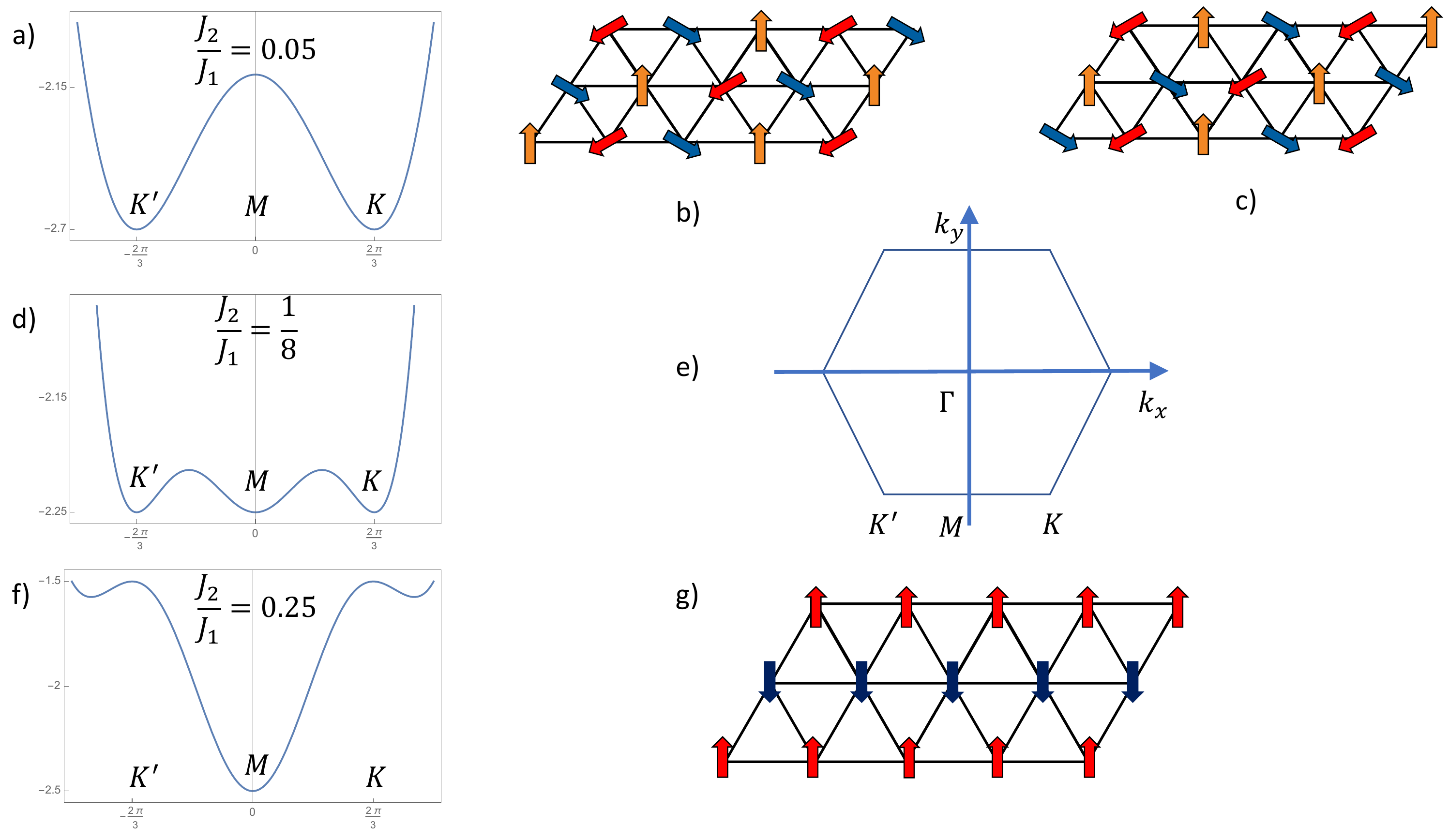}}
	\caption{(Color online) (a), (d), (f): Single particle dispersion relations on a triangular lattice for different values of $t=J_2/J_1$. Schematic nature of the phases condensed at the $K$ (b), $K'$ (c) and $M$ (g)  points of the BZ (e). 
	}
	\label{fig01}
\end{figure}

The Hamiltonian~(\ref{H-XXZ}) can be regarded
as a model of hard-core bosons hopping on a triangular lattice with NN  amplitude $J_1$ and NNN amplitude $J_2$. At small $J_2$ (small $t<1/8$), the single boson dispersion  exhibits two degenerate minima located at the $K$ and $K^\prime$ points of the Brillouin zone (BZ), Fig.~\ref{fig01}a. This implies that noninteracting bosons can  condense to any superposition of these two states, however the hard-core interactions prevent forming a density modulation and enforce condensation into one of these two points. This leads to the doubly degenerate ground states, identified with the planar $120^\circ$ N\'{e}el configurations of spins with two helicities, Figs.~\ref{fig01}b,c. 
At $t=1/9$ the single particle dispersion acquires an additional minimum at the $M$ point, midway between $K$ and $K^\prime$, while at  $t=1/8$ the dispersion is triply degenerate, Figs.~\ref{fig01}d and \ref{fig01}e. At $t>1/8$ the global minimum is at $M$, signaling  semiclassically a first order transition\cite{Chubukov-17,Chubukov-18} into a state with the collinear stripe order  shown in Fig.~\ref{fig01}g.

\section{Emergence of anisotropic Dirac fermions} 
This  picture is severely modified by quantum fluctuations. We account 
for these by reformulating the model (\ref{H-XXZ}) as a
theory of {\em spinless} lattice fermions coupled to a CS gauge field. The fermionization automatically takes care of the hard-core condition.  The spin operators may be represented as
%\bea
%\label{CS-fermion}
$ S^{\pm}_{\bf r}=\exp \left( i e \sum_{{\bf r'}\neq{\bf r}}\arg[{\bf r}-{\bf r'}]n_{\bf r'}\right)f^{\pm}_{\bf r}, $
%\ena
where $e=2l+1$ is an odd integer representing the CS charge, $f^{\pm}_{\bf r}$ are  creation/annihilation operators of canonical spinless fermions,
%obeying canonical anticommutation relations, 
$n_{\bf{r}}=f^\dagger_{\bf r} f_{\bf r}=S^+_{\bf r}S^-_{\bf r}$ is the particle number operator,
and summation runs over all lattice sites.  The XX part of the Hamiltonian (\ref{H-XXZ}) acquires the form: 
 \bea
\label{H2}
H_l=  \frac{J_l}{2} \sum_{{\bf r},\nu}f^\dagger_{\bf r}f_{{\bf r}+{\bm \mu}_\nu^l}e^{i \Lambda_{{\bf r}, {\bf r}+{\bm \mu}^l_\nu} }+H.c.,  
\eea
where ${ \Lambda}_{{{\bf r}_1}, {\bf r}_2}=\sum_{\bf r}\left[\arg({\bf r}_1-{\bf r})-\arg({\bf r}_2-{\bf r})\right] n_{\bf r}$ is a  
gauge field associated with the NN and NNN links on the triangular lattice.
It introduces CS magnetic flux threading the unit cell of the triangular lattice:  ${\Phi}_{\bf r}={\Lambda}_{{\bf r},{\bf r}+{{\bf e}_1}}+{\Lambda}_{{\bf r}+{\bf e}_1,{\bf r}+{{\bf e}_1}+{\bf e}_2}
+{\Lambda}_{{\bf r}+{{\bf e}_1}+{\bf e}_2,{\bf r}+{\bf e}_2}+{\Lambda}_{{\bf r}+{{\bf e}_2},{\bf r}}$, which is the lattice analog of  ${\Phi}_{\bf r}=\curl {\Lambda}$ (for details see Appendix A).
The Hamiltonian~(\ref{H-XXZ}) thus can be rewritten in terms of fermions \cite{SKG2,SKG,SGK1,maiti,CSS,WWS}, $f_{\bf r}$, coupled to the $U(1)$ CS gauge field \cite{SKG2}.

\begin{figure}[t]
	\center
	%\scalebox{0.5}{\rotatebox{0}{\includegraphics{Mbox.eps}}}
	\centerline{\includegraphics[width=75mm,angle=0,clip]{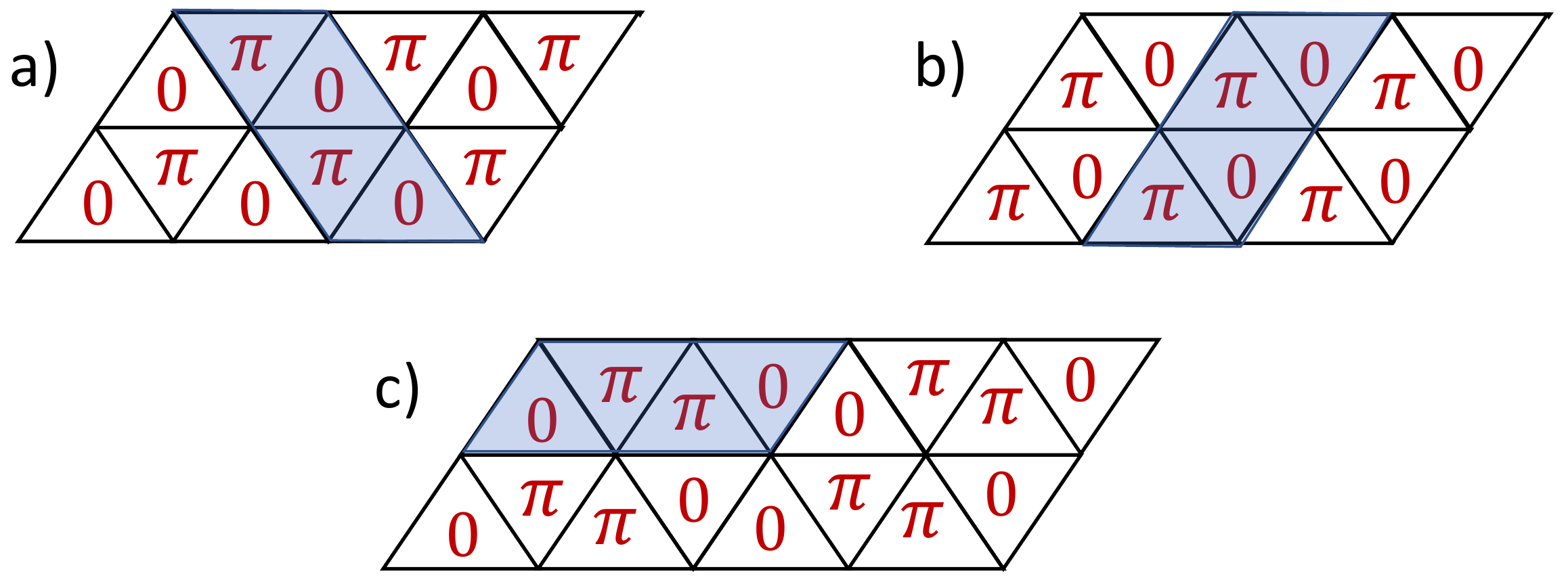}}
	\caption{(Color online) a) and b) Distinct $\pi$ flux configurations of CS fermionization,  corresponding to two helicities of the $120^\circ$ order. 
	%c) $\pi$ flux configuration at finite $J_2$ corresponding to a NNN triangular sub-lattice. Each shaded $120^\circ$ triangle is threaded by $\pi$ flux. 
	c) $\pi$ flux configuration corresponding to the stripe order.}
	\label{fig02}
\end{figure}

To illustrate how this  $U(1)$ field affects the fermion dynamics,
we analyze the XX limit of the Hamiltonian~(\ref{H-XXZ}). 
%{\em Chern-Simons superconductor description of the $120^\circ$ ordered state}. - 
In the absence of a net magnetization the CS fermion state is half-filled, $\langle n \rangle =1/2$.  
This implies that the CS phases create ${2 i \pi \langle n \rangle}\rightarrow i\pi$ flux,  threading the unit cell.
The double degeneracy (two helicities) of the $120^\circ$ state  is reflected in the staggered $\pi$ flux patterns, Figs.~\ref{fig02}a,b, for which there are two inequivalent choices,
distinguished by the sign of the $z$-component of the 
{\em vector chirality}, defined on a triangular plaquette  as\cite{villain, KW}
\begin{equation}
					\label{eq:vector-chirality}
\kappa_z=\epsilon_{ij}\left(\langle \hat{S}^i_1 \hat{S}^j_2\rangle +\langle S_2^i \hat{S}_3^j\rangle + \langle \hat{S}_3^i\hat{S}^j_1\rangle\right),
\end{equation} 
where $\langle\ldots\rangle$ stands for the quantum expectation value and $i,j=x,y$.
Importantly, this $\mathbb{Z}_2$ order parameter, reflecting the two different  $\pi$ flux patterns, persists in the SL phase, again implying a double degeneracy. %Each of these patterns leads to the doubling of the unit cell.

\begin{figure}[h]
	\center
	%\scalebox{0.5}{\rotatebox{0}{\includegraphics{Mbox.eps}}}
	\centerline{\includegraphics[width=75mm,angle=0,clip]{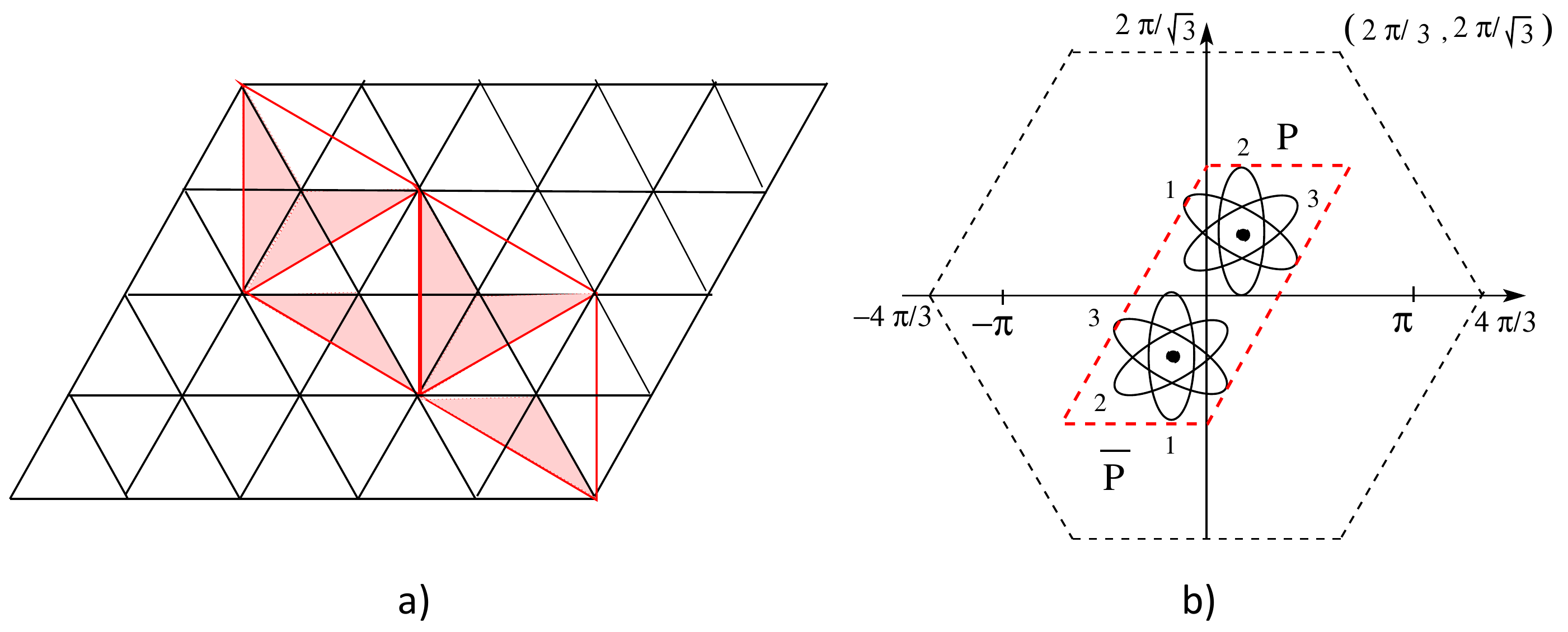}}
	\caption{a) $\pi$ flux configuration on the NNN large triangular sub-lattice, $\tau=1$. Each shaded $120^\circ$ triangle is threaded by $\pi$ flux. 
b)	Reduced BZ (red), $1/6^{th}$ of the original BZ (black), with Dirac points $P$ and $\bar P$. The ellipses indicate anisotropic dispersion relations of Dirac fermions labeled by $\tau=1,2,3$\ .	%The reduced BZis formed by a rhombus that includes a pair of Dirac points $(P_\tau, \bar{P}_\tau)$. 
} 
	\label{fig011}
\end{figure}

Consider now the fermionized Hamiltonian for  $J_2>0$. The NNN bonds form three disjoint large triangular lattices, labelled as $\tau=1,2,3$. At half-filling, fermions hopping along these  bonds still accumulate a $\pi$ flux through a large rhomboidal cell composed of NNN links. The CS transformation unambiguously identifies the $\pi$-flux  configuration depicted in Fig.~\ref{fig011}a for one of the three sub-lattices, $\tau=1$.
Flux patterns of the two other NNN sub-lattices are obtained by rotation of Fig.~\ref{fig011}a by $\pm 2\pi/3$. Such an arrangement of $\pi$ fluxes on three NNN triangular lattices preserves  $C_3$ symmetry. 
However, it further violates the translational invariance in three lattice spacings
and, thus the unit cell becomes 6 times larger than the original one. Correspondingly,
there is a 6 component Fermi field, $f^\tau_{{\mathbf k},\alpha}$, where $\alpha=1,2$ and $\tau=1,2,3$,  in a 6 times reduced BZ, leading to a $6 \times 6$ Hamiltonian (for details see Appendix B).

The corresponding band structure exhibits two Dirac points, $P$ and $\bar{P}$, in the reduced BZ,  Fig~\ref{fig011}b.
The momentum space Hamiltonian expanded near the Dirac points  is 
${\mathcal H}^L=H_{0}^{(P)}+H_{0}^{(\bar{P})} + H_{\mathrm int}$, where
\bea
                                                                    \label{Dirac-1}
\!\!\!\!\!\!H_{0}^{(P)}({\bf k})= v \sum_{{\bf k,\nu,\tau}}\hat{f}^{\tau\dagger}_{\bf k,\alpha}\big[(p_\nu+t q_\nu) \sigma^\nu_{\alpha\beta}
 -2t q_\tau\sigma^\tau_{\alpha\beta}
	 \big] \hat{f}^\tau_{\bf k,\beta},
\ena
and its time-reversed partners $H_{0}^{(\bar{P}_\tau)}({\bf k})=\left[H_{0}^{(P_\tau)}(-{\bf k})\right]^*$ with 
$\hat{f}^\tau_{ \bf k, \alpha} \rightarrow {\hat{\bar{f}}^\tau}_{\!\!\!\!\! \bf k, \alpha}$,
and  $v=J_1a$,  $a$  being the lattice constant. We use: $p_\nu= {\bf k}\cdot {\bf e}_\nu$ and  $q_\nu= {\bf k}\cdot{\bf a}_\nu, \;\nu=1,2,3$. Hereafter 
we work with dimensionless quantities measuring  energy in units of $J_1$ and momentum  in units of  $1/a$. 
  The low-energy fermion operators, 
$\hat{{f}^\tau}_{\!\!\!\!\! \bf k, \alpha}$ and  ${\hat{\bar{f}}^\tau}_{\!\!\!\!\! \bf k, \alpha}$ 
have momenta measured from  $P$ and $\bar{P}$ points, respectively.  
  The $C_3$ invariance is ensured by $2\pi/3$ rotations of the lattice accompanied with cyclic transformation, $\tau \rightarrow \tau+1$,  of the fermion copy.
  The Hamiltonian (\ref{Dirac-1}) leads to the {\em anisotropic} spectrum 
  \bea
  \label{spec-1}
 E^\tau_{0,\bf k}=\pm \Big[(1+ 3 t^2 )\sum_{\nu=1}^3 p^2_\nu  -4 t p_\tau q_\tau \Big]^{1/2}.
 % \; \tau =1,2,3 
  \ena  
 The $120^\circ$ state with the other helicity corresponds to ${\mathcal H}^R=H_{0}^{(P^\prime)}+H_{0}^{(\bar{P}^\prime)} + H_{\mathrm int}$. Note that the ground state spontaneously chooses one of the two.% helicities/$\pi$-flux configurations (Fig.~2a,b).  

Upon a gauge transformation, the CS phases in Eq.~(\ref{H2}) may be rewritten as covariant derivatives, leading to the substitution ${\bf k} \to {\bf k} -e{\bf A}_{\bf r}$ in Eq.~(\ref{Dirac-1}). Here ${\bf A}_{\bf r}$ in the kinetic term reflects fluctuations of  the CS phases from $0$ or $\pi$ per plaquette and is bilinear in fermion operators. It thus generates a two particle interaction vertex \cite{CSS,WWS}
\bea
\label{Dirac-2}
H_{\mathrm int}= - \!\!\!\sum_{\bf k,k',q,\tau}\!\!\! V^{\alpha\alpha',\beta\beta'}_{\bf q} \hat{f}^{\tau\dagger}_{{\bf k},\alpha}\hat{\bar{f}}^{\tau\dagger}_{{\bf k'+q},\alpha'}
\hat{\bar{f}}^\tau_{{\bf k'},\beta} \hat{f}^\tau_{{\bf k+q},\beta'},
\ena
where 
$V^{\alpha\alpha',\beta\beta'}_{\bf q}=2 \pi i e  \left(\sigma^\nu_{\alpha\beta}\delta_{\alpha'\beta'}+
\delta_{\alpha\beta}[\sigma^{\nu}]^T_{\alpha'\beta'}\right) B^\nu_{\bf k}$
where   $\nu=1,2,3$ and  $B^\nu_{\bf k}= \epsilon_{ij} A^j_{\bf k} (e^i_\nu + t a^i_\nu) $ is  determined by
the Fourier image ${\bf A}_{\bf k}={\bf k}/|{\bf k}|^2$ of the  vector potential of the vortex gauge field 
${\bf A}_{\bf r}$ defined above.

\section{Chern-Simons superconductivity: description of phases and deconfined phase transition}

In this section, we will discuss the Chern-Simons superconductor description of ordered phases, the emergence of the Helical spin-liquid, and the deconfined phase transition from $120^\circ$ state to the Helical spin-liquid.

\subsection{Chern-Simons superconductor description of the $120^\circ$ state}

The CS  interaction (\ref{Dirac-2}) leads to the Cooper pairing of fermions residing near the $P$ and $\bar P$ points and may result in a  broken U(1) superconducting phase. In terms of the original spins the latter corresponds to a broken O(2)  $120^\circ$ antiferromagnet. Upon increasing $t$, the fermion dispersion becomes more anisotropic, weakening the Cooper pairing (which operates only within time-reversal pairs with the same NNN sub-lattice index $\tau$, Eq.~(\ref{Dirac-2}), which has a non-collinear anisotropy orientation in $P$ and $\bar P$  points, Fig.~\ref{fig011}b). This leads to an eventual  breakdown of the CS superconductivity at $t\approx 0.089$.    

\begin{figure}[t]
 	\center
 	%\scalebox{0.5}{\rotatebox{0}{\includegraphics{Mbox.eps}}}
 	\centerline{\includegraphics[width=65mm,angle=0,clip]{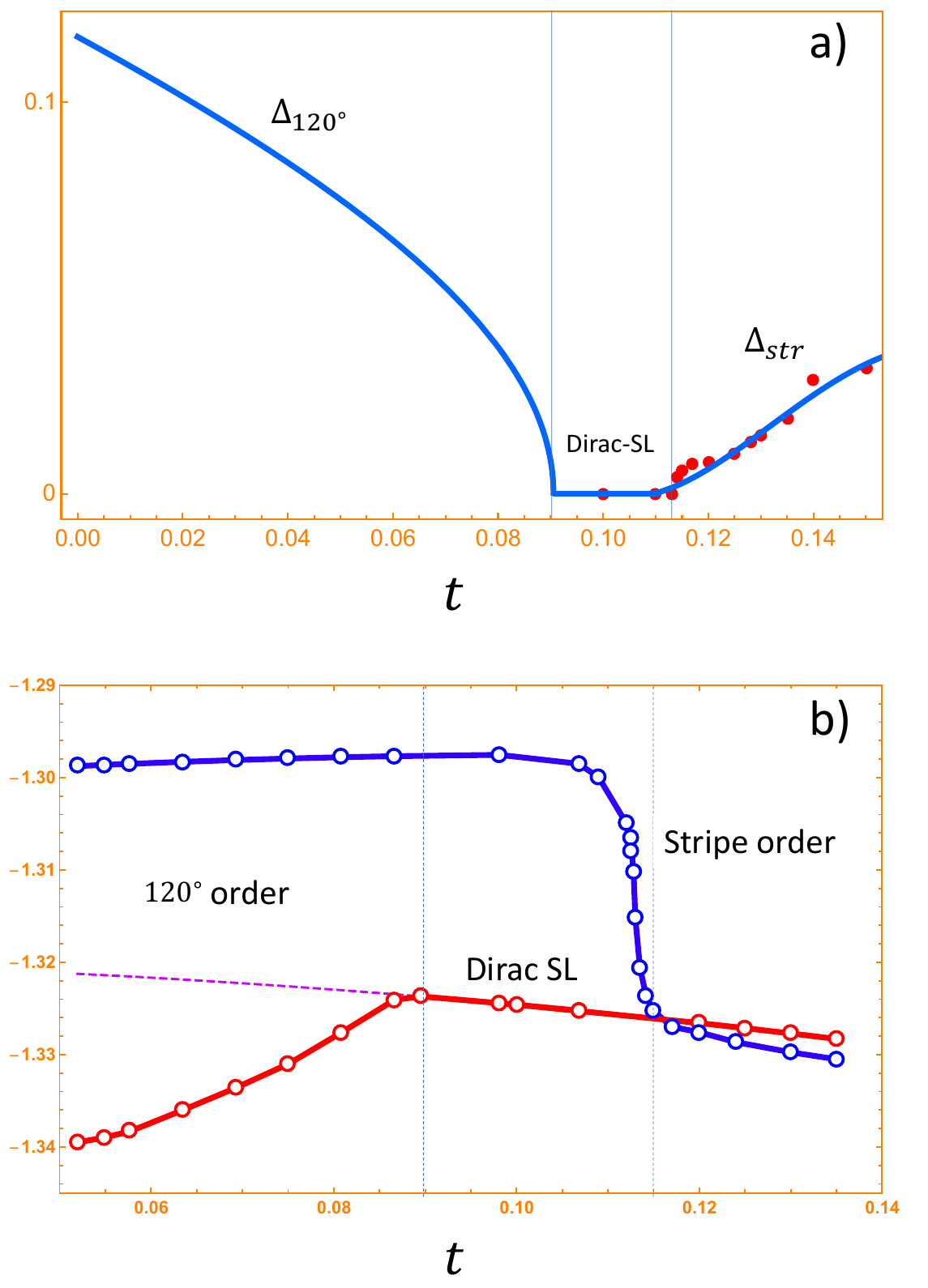}}
 	\caption{(Color online) (a) Excitation gap of the fermion spectrum plotted versus $t=J_2/J_1$ in the CS superconductor mean-field description of the $120^\circ$ state, 
	Dirac SL, and the CS superconductor mean-field description of the striped phase. 
	The superconducting order parameters are zero at $0.089 \lesssim t\lesssim 0.113$.
	Note that the order parameter $\Delta_{120^\circ}$ vanishes continuously near $t\sim 0.089$, signaling a continuous phase transition to the SL phase. (b) Ground state energy of gapless Dirac fermions on a triangular lattice at half-filling with $\pi$-flux distribution corresponding to Fig.~(2a) and/or (2b) as a function of $t$ compared with the energy of the  CS superconductor of the striped phase. Note a first order transition at the level crossing near $t\sim 0.116$. }
 	\label{fig04}
 \end{figure}

To describe this physics we employ  the standard BCS treatment with the superconducting order parameter $\Delta^{\alpha\alpha'}_{\bf k}=-2\pi i e 
	\sum_{\beta\beta' \bf k'}V^{\alpha\alpha',\beta\beta'}_{\bf k-k'}\langle\hat{\bar{f}}_{ -\bf k', \beta} \hat f_{ \bf k',\beta'}\rangle$,
where the index $\tau$ is dropped hereafter. The order parameter is quadratically coupled to the fermions as: $\sum_{\alpha\alpha' \bf k}\Delta^{\alpha\alpha'}_{\bf k} \hat{f}^\dagger_{ \bf k, \alpha} \hat{\bar f}^\dagger_{ -\bf k,\alpha'} + H.c.$, leading to self-consistency conditions. 
Following Ref.~[\onlinecite{CSS}], one expects   $p\pm ip$ symmetry of $\Delta^{\alpha\alpha'}_{\bf k}$ and thus looks for the solution  in the form
	\bea
	\label{p}
	\Delta^{\alpha\alpha'}_{\bf k} =\Delta_{3 \bf k}\delta_{\alpha\alpha'}
	+i \frac{\Delta_{0,\bf k}}{\sqrt{3}}\!\! \sum_{\mu,\nu=1,2,3}\!\!\!\eta_{\mu\nu}(q_\mu-3 t p_\mu)
	\sigma^\nu_{\alpha\alpha'},
	\ena 
where $\eta_{\mu\nu}$ is defined as $\eta_{33}=1, \eta_{ij}=\epsilon_{ij}, \eta_{3i}=0,\; i,j=1,2$. The corresponding self-consistency equations in terms of the scalar order parameters $\Delta_{0,\bf k}$ and $\Delta_{3,\bf k}$ are given in 
Appendix C. At $t=0$ they exhibit a non-trivial solution~\cite{CSS} for CS charge $e=3$. 
For $t>0$ the anisotropy of the Dirac spectrum (\ref{spec-1}), Fig.~\ref{fig011}b, suppresses the  order parameter.   Fig.~\ref{fig04}a shows the corresponding gap in the fermionic spectrum, $\Delta_{120^\circ}$, obtained through a numerical solution of the self-consistency equations. The gap and the U(1) broken state collapse at $t=0.089$, indicating the absence of the planar long range order at larger $t$. Notice that the $\mathbb{Z}_2$ symmetry breaking, associated with the choice of CS flux pattern, Fig.~\ref{fig02}a,b, remains intact across this transition.

\subsection{Helical spin-liquid phase}
For $t>0.089$, in the CS superconductivity associated with one of the two $120^\circ$ states, one is left thus with a gapless state with an unbroken U(1) symmetry and excitations described in terms of $N=6$ copies of Dirac fermions with the anisotropic dispersion. This is the SL ground state,  doubly degenerate due to the presence of the the long-range $\mathbb{Z}_2$ order.  The latter may be detected by a finite value of the 
vector chirality, Eq.~(\ref{eq:vector-chirality}).  To derive an effective  low-energy field theory in this regime 
one should integrate out fermionic degrees of freedom with momenta away from  the two Dirac points. This way one obtains a stable\cite{hermele} 2+1 dimensional Maxwell electrodynamics coupled to $N=6$ copies of the anisotropic Dirac fermions (see Appendix D).  

An external magnetic field in the $z$ direction, $H_h=h\sum_{\mathbf r} S_{\mathbf r}^z$, leads to a deviation from half filling and thus to a non-zero chemical potential of the Dirac fermions.  Each additional fermion comes with an extra flux quantum of the gauge field. It results in Landau quantization of the fermionic energies with the fully filled levels. Thus the excitation spectrum of the SL in a magnetic field is {\em gapped} with the gap proportional to $|h|$.

\subsection{Chern-Simons superconductor description of the stripe phase}
Consider an alternative choice of CS flux pattern, depicted in Fig.~\ref{fig02}c. This choice breaks the $C_3$ symmetry and selects a preferred direction along the lattice. 
The reduced BZ can be chosen to be the same as in Fig.~\ref{fig011}b,    but the Dirac points are $Q^{\pm}=(\pm \arccos(t),0)$. 
The superconducting solution (see Appendix C) exists for $t\gtrsim 0.113$. However its energy is smaller than that of SL only at $t\gtrsim 0.116$, Fig.~\ref{fig04}b, suggesting a first order transition. 
The corresponding U(1) broken state is the stripe phase, Fig.~\ref{fig01}g, which also breaks $C_3$ lattice symmetry. 
As a result, the helical SL state appears stabilized in the narrow interval $0.089\lesssim t \lesssim 0.116$.

\section{Discussion and estimates}
There are two main differences between our work and previous approaches to establish spin-liquids: in our framework, we start by focusing on the ordered states of the spin-1/2 XXZ magnet via treating it as superconducting states of spinless CS fermions. This approach is as good as other methods (e.g., using Schwinger or Holstein-Primakoff representations of spins) to describe the symmetry broken state. We then study the stability region of the ordered state, taking into account that in the spin-liquid phase the ordering breaks down. 
%We formulated a framework where the ordered states of the XXZ magnet are treated as superconducting states of spinless CS fermions. 
Thus, the SL emerges in the parameter window where such superconducting 
states cannot be established. It would be desirable to test these predictions with, say, DMRG simuations, although the large number of Dirac points involved may make this technically challenging. 

We have mostly focused on the XX case, $g=0$, while a finite $g$ leads to an additional fermionic interaction vertex, which modifies the self-consistency equations. This leads in turn to a weak $g$-dependence of the critical values of $t$. Since the model does not have SU(2) symmetry, the low-energy fermion excitations may be considered spinless, and as such, the SL is outside the projective symmetry group classification of SL's based on Schwinger boson and Abrikosov fermion representation of spins \cite{AV,XGV,Bieri}. We encourage neutron scattering experiments for observation of the proposed spin liquid state. Here the spin magnetization distribution and spin-spin correlators are expected not to exhibit Bragg peaks. 
Moreover, the anisotropy of the dynamical structure factor near Dirac points may serve as a signal of the anisotropic Dirac dispersion.
The information about the vector chiral order of the spin-liquid may be revealed in nuclear-magnetic interferences in the chiral magnetic scattering\cite{Maleyev,Simonet,ER} of an initially unpolarized neutron beam.

\section*{Acknowledgement}
We are indebted to L. Balents, A. Chernyshev, A. Chubukov, O. Starykh and Mengxing Ye for valuable discussions. A.K. was supported by NSF grant DMR-1608238. 
T.A.S. acknowledges startup funds from UMass Amherst and thanks the Max Planck Institute for the Physics of Complex Systems for hospitality.

%%%%%%%%%%%%%%%
\appendix

\begin{widetext}

\section{Flux configuration for Chern-Simons fermions on the triangular lattice}

The Chern-Simons (CS) transformation 
\begin{eqnarray}
 \label{transform}
 S^{\pm}_{\bf r}=\exp \left( i e \sum_{{\bf r'}\neq{\bf r}}\arg[{\bf r}-{\bf r'}]n_{\bf r'}\right)f^{\pm}_{\bf r}, 
 \end{eqnarray} 
where the summation is performed over all lattice sites except $r$,
defines the distribution of position-dependent  phases on links of the lattice (see Eq.~(2) of the main text for definitions). The phase corresponding to the link between sites positioned at ${\bf r}_1$ and ${\bf r}_2$ can be divided into two parts:
\bea
\label{phase-0}
\Lambda_{{\bf r}_1,{\bf r}_2}=\sum_{{\bf r'}\neq{\bf r}_1}\arg[{\bf r}_1-{\bf r'}]n_{\bf r'}-\sum_{{\bf r'}\neq{\bf r}_2}\arg[{\bf r}_2-{\bf r'}]n_{\bf r'}=\Lambda^\Gamma_{{\bf r}_1,{\bf r}_2}+\Lambda^L_{{\bf r}_1,{\bf r}_2}.
\ena
The first term, $\Lambda^\Gamma_{{\bf r}_1,{\bf r}_2}$, is the "global" phase formed from all points ${\bf r}$ away from the link $({\bf r}_1,{\bf r}_2)$,
\bea
\label{phase-G}
\Lambda^\Gamma_{{\bf r}_1,{\bf r}_2}=\sum_{{\bf r'}\neq{\bf r}_1,{\bf r}_2}\Big[\arg[{\bf r}_1-{\bf r'}]-\arg[{\bf r}_2-{\bf r'}]\Big]n_{\bf r'},
\ena
accounting for the scanning angles on the lattice outside of the link ${\bf r}_1,{\bf r}_2$.
The second term, $\Lambda^L_{{\bf r}_1,{\bf r}_2}$, is the "local" phase formed from the end points of links
\bea
\label{ep}
\Lambda^L_{{\bf r}_1,{\bf r}_2}=\arg({\bf r}_2-{\bf r}_1)n_{{\bf r}_1}-\arg({\bf r}_1-{\bf r}_2)n_{{\bf r}_2}.
\ena

Now we consider the unit cell rectangle presented in Fig.\ref{sup-fig-1}(left panel) and calculate the phase acquired by a fermion upon clockwise rotation around it (this is a convention we use). 
First, we calculate the net phase accumulated from local phases $\Lambda^L_{{\bf r}_i,{\bf r}_j}$, coming from the endpoints of the links.  According to this expression, one can find the following phases on links depicted in Fig.\ref{sup-fig-1}(left panel):
\bea
\label{phases}
\Lambda^L_{{\bf r}-{\bf e}_1,{\bf r}-{\bf e}_1+{\bf e}_2}&=&\pi n_{{\bf r}-{\bf e}_1+{\bf e}_2}, \;\;\;\;\; \Lambda^L_{{\bf r}-{\bf e}_1+{\bf e}_2,{\bf r}+{\bf e}_2}=-\frac{\pi}{3} n_{{\bf r}-{\bf e}_1+{\bf e}_2}+\frac{4 \pi}{3} n_{{\bf r}+{\bf e}_2},\;\;\;\;
\Lambda^L_{{\bf r}+{\bf e}_2,{\bf r}}=-\pi n_{{\bf r}+{\bf e}_2}, \nn\\
\Lambda^L_{{\bf r},{\bf r}-{\bf e}_1}&=&\frac{\pi}{3} n_{{\bf r}-{\bf e}_1}-\frac{4 \pi}{3} n_{\bf r},\;\;\;
\Lambda^L_{{\bf r}-{\bf e}_1+{\bf e}_2,{\bf r}}=-\frac{2\pi}{3} n_{{\bf r}-{\bf e}_1+{\bf e}_2}+\frac{5 \pi}{3} n_{\bf r}
\ena
Here the convention for $\arg$ functions is adopted and they are calculated with respect to the  $x$-axes in a counterclockwise direction.
%%%%%%%%%%%%%%
\begin{figure}[h]
	\centerline{\includegraphics[width=150mm,angle=0,clip]
		{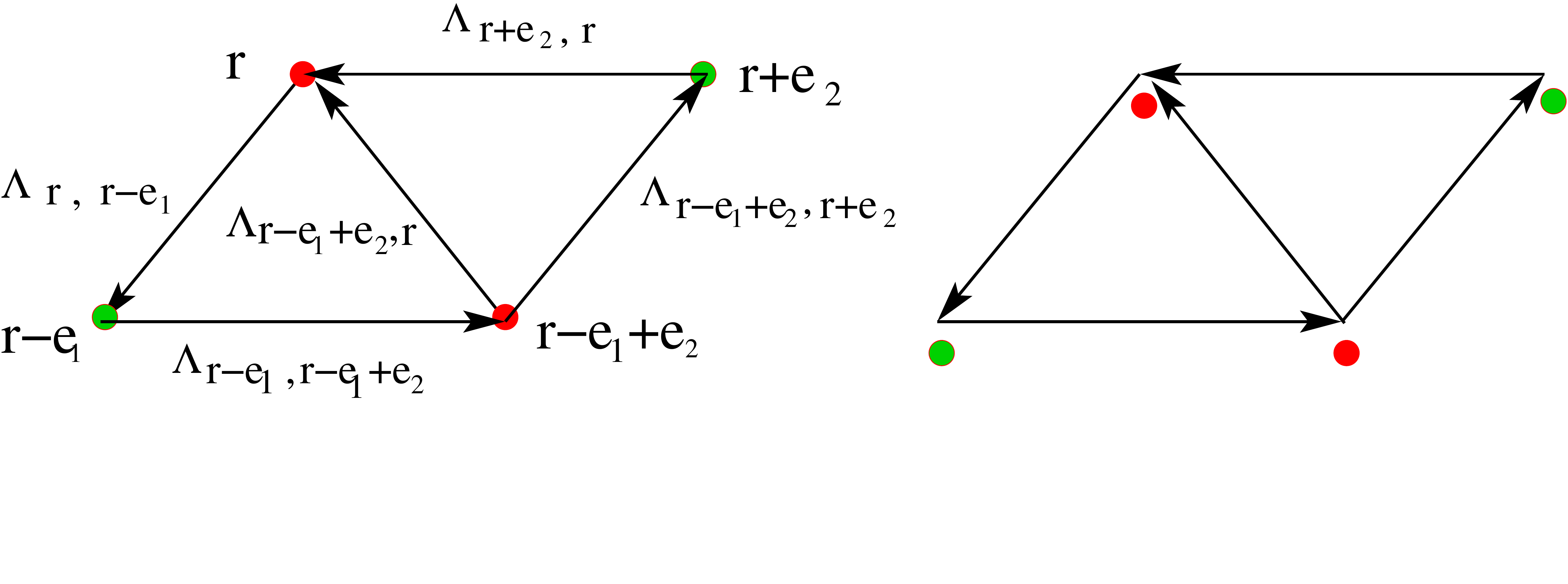}}
	\caption{(Color online) Left panel: A unit cell of the triangular lattice. Phases associated with the links are shown. Right panel: The unit cell is "remedied" with an infinitesimal shift showing 
that the cell includes one site.} 
	\label{sup-fig-1}
\end{figure}
%%%%%%%%%%%%
The accumulated local fluxes over triangles in the unit cell will be
\bea
\label{phases-2}
\Phi^L_{{\bf r}-{\bf e}_1,{\bf r}-{\bf e}_1+{\bf e}_2 ,{\bf r}}&=&\Lambda^L_{{\bf r}-{\bf e}_1,{\bf r}-{\bf e}_1+{\bf e}_2 }+\Lambda^L_{{\bf r}-{\bf e}_1+{\bf e}_2 ,{\bf r}}+\Lambda^L_{{\bf r},{\bf r}-{\bf e}_1}=\frac{\pi}{3}(n_{{\bf r}-{\bf e}_1}+n_{{\bf r}-{\bf e}_1+{\bf e}_2}+n_{\bf r}),\\
\Phi^L_{{\bf r}-{\bf e}_1+{\bf e}_2 ,{\bf r}+{\bf e}_2,{\bf r}}&=&\Lambda^L_{{\bf r}-{\bf e}_1+{\bf e}_2 ,{\bf r}+{\bf e}_2}+\Lambda^L_{{\bf r}+{\bf e}_2,{\bf r} }+\Lambda^L_{{\bf r},{\bf r}-{\bf e}_1+{\bf e}_2}
=\frac{\pi}{3}(n_{{\bf r}-{\bf e}_1+{\bf e}_2}+n_{{\bf r}+{\bf e}_2}+n_{\bf r})-2 \pi n_{\bf r} \nn.
\ena
The local flux corresponding to the unit cell will thus be:
\bea
\label{phase-3}
\Phi^L_{{\bf r},{\bf r}+{\bf e}_1 ,{\bf r}+{\bf e}_1+{\bf e}_2,{\bf r}+{\bf e}_2}&=&\frac{ \pi}{3}(n_{{\bf r}-{\bf e}_1}+n_{{{\bf r}+{\bf e}_2}}+2 n_{\bf r}+2 n_{{{\bf r}-{\bf e}_1+{\bf e}_2}})-2 \pi n_{\bf r}.
\ena

The calculation of global fluxes through triangles is
much simpler. The phase on a given link in a triangle is equal to the opposite 
to the link angle times the density operator corresponding to that site. It is straightforward to obtain
 \bea
 \label{phases-4}
 \Phi^\Gamma_{{\bf r}-{\bf e}_1,{\bf r}-{\bf e}_1+{\bf e}_2 ,{\bf r}}&=&\Lambda^\Gamma_{{\bf r}-{\bf e}_1,{\bf r}-{\bf e}_1+{\bf e}_2 }+\Lambda^\Gamma_{{\bf r}-{\bf e}_1+{\bf e}_2 ,{\bf r}}+\Lambda^\Gamma_{{\bf r},{\bf r}-{\bf e}_1}=\frac{\pi}{3}(n_{{\bf r}-{\bf e}_1}+n_{{\bf r}-{\bf e}_1+{\bf e}_2}+n_{\bf r}),\\
 \Phi^\Gamma_{{\bf r}-{\bf e}_1+{\bf e}_2 ,{\bf r}+{\bf e}_2,{\bf r}}&=&\Lambda^\Gamma_{{\bf r}-{\bf e}_1+{\bf e}_2 ,{\bf r}+{\bf e}_2}+\Lambda^\Gamma_{{\bf r}+{\bf e}_2,{\bf r} }+\Lambda^\Gamma_{{\bf r},{\bf r}-{\bf e}_1+{\bf e}_2}=\frac{\pi}{3}(n_{{\bf r}-{\bf e}_1+{\bf e}_2}+n_{{\bf r}+{\bf e}_2}+n_{\bf r}) \nn.
 \ena
The global flux through the rectangular unit cell will be
 \bea
 \label{phase-5}
 \Phi^\Gamma_{{\bf r},{\bf r}+{\bf e}_1 ,{\bf r}+{\bf e}_1+{\bf e}_2,{\bf r}+{\bf e}_2}&=&\frac{ \pi}{3}(n_{{\bf r}-{\bf e}_1}+n_{{{\bf r}+{\bf e}_2}}+2 n_{\bf r}+2 n_{{{\bf r}-{\bf e}_1+{\bf e}_2}}).
 \ena 
 
 Finally, one can make use of the obtained expressions to calculate the the total flux threading each of the triangles of the unite cell and the total flux through the the rectangular unit cell itself. 
 Putting all together we see that 
 \bea
 \label{phase-6}
 \Phi_{{\bf r}-{\bf e}_1,{\bf r}-{\bf e}_1+{\bf e}_2 ,{\bf r}}&=& \Phi^\Gamma_{{\bf r}-{\bf e}_1,{\bf r}-{\bf e}_1+{\bf e}_2 ,{\bf r}}+ \Phi^L_{{\bf r}-{\bf e}_1,{\bf r}-{\bf e}_1+{\bf e}_2 ,{\bf r}}\nn\\
 &=& \frac{2\pi}{3}(n_{{\bf r}-{\bf e}_1}+n_{{\bf r}-{\bf e}_1+{\bf e}_2}+n_{\bf r}),\\
 \Phi_{{\bf r}-{\bf e}_1+{\bf e}_2 ,{\bf r}+{\bf e}_2,{\bf r}}&=& \Phi^\Gamma_{{\bf r}-{\bf e}_1+{\bf e}_2 ,{\bf r}+{\bf e}_2,{\bf r}}+ \Phi^L_{{\bf r}-{\bf e}_1+{\bf e}_2 ,{\bf r}+{\bf e}_2,{\bf r}}\nn\\
 &=&\frac{2\pi}{3}(n_{{\bf r}-{\bf e}_1+{\bf e}_2}+n_{{\bf r}+{\bf e}_2}+n_{\bf r})-2 \pi n_{\bf r},\\ 
 \Phi_{{\bf r},{\bf r}+{\bf e}_1 ,{\bf r}+{\bf e}_1+{\bf e}_2,{\bf r}+{\bf e}_2}&=&\Phi^\Gamma_{{\bf r},{\bf r}+{\bf e}_1 ,{\bf r}+{\bf e}_1+{\bf e}_2,{\bf r}+{\bf e}_2}+\Phi^L_{{\bf r},{\bf r}+{\bf e}_1 ,{\bf r}+{\bf e}_1+{\bf e}_2,{\bf r}+{\bf e}_2}\nn\\
&=&\frac{2 \pi}{3}(n_{{\bf r}-{\bf e}_1}+n_{{{\bf r}+{\bf e}_2}}+2 n_{\bf r}+2 n_{{{\bf r}-{\bf e}_1+{\bf e}_2}})-2 \pi n_{\bf r}.
 \ena
Here we see that the uniform density smearing approximation, $n_{\bf r}\rightarrow n=const$ ($n$ being the lattice filling fraction), still implies that fluxes threading the triangles of the unit cell are not uniform but are modulated: The flux through one of the regular triangles in the unit cell is vanishing while the flux threading the other regular triangle becomes $\Phi\rightarrow 2\pi n$. Such a flux modulation can be qualitatively understood as follows.  
 Fig.~\ref{sup-fig-1}(right panel) depicts the unit cell, which is shown with an infinitesimally small shift of the sites. Depending on the direction of the shift, the site corresponding to the unit cell is located within one of the two regular triangles. Hence the net flux $2\pi n$ is threading only one of the two regular triangles. At half filling $n=1/2$ and thus one obtains a $\pi$-flux lattice, as shown in Fig.3 of the main text.

\section{Hamiltonian}
The CS transformation \cite{SKG2} implies a staggered $\pi$ flux distribution within the NN
and NNN triangular sub-lattices.  
The flux distribution in the NNN triangular sublattices with $\tau=2,3$ is obtained from the lattice corresponding to $\tau=1$ upon rotation
by $\pm2\pi/3$, respectively. 
Such a phase distribution breaks translational
invariance on three lattice steps reducing the Brillouin zone of the NN triangular lattice 6 times.

 As a result of the staggered $\pi$-flux threading of every other triangle in the unit cell (including both triangles composed on NN and NNN bonds), 
 the single-fermion dispersion on a triangular lattice will acquire a Dirac form around the following points of the first Brillouin zone:
 $P_\tau =(\pi/6+ 2\pi \tau/3, \pi/2\sqrt{3})$ and $\bar{P}_\tau =(-\pi/6+ 2\pi \tau/3, \pi/2\sqrt{3})$, $\tau=1,2,3$, giving rise to six components of the Fermi field.
These Dirac points form a triangular lattice in momentum space (see Fig.~4b of the main text), while the reduced BZ is a rhombus that includes only
one pair of points $(P_\tau, \bar{P}_\tau)$.

The double degeneracy of the planar $120^\circ$ state  is linked to the interchange of fluxes $\pi\leftrightarrow 0$ threading each triangular face of the unit cell. 
We note that the chiralities of the Hamiltonian expanded around $P_\tau$ and $\bar{P}_\tau$ are opposite. Similarly, if one identifies the refletion of $P_\tau$ and $\bar{P}_\tau$
with respect to the $k_x$-axis  with $P'_\tau$
and $\bar{P}'_\tau$, then the chiralities of the corresponding Hamiltonians (expanded around $P'_\tau$
and $\bar{P}'_\tau$) will also be opposite to each other. 
 This implies that the single particle states at $t\rightarrow 0$, in close vicinities of these Dirac points, are given by 
$\bar{u}_\tau^\prime({\bf k})=u_\tau({\bf k})=\frac{1}{\sqrt{2}}\left(
\begin{array}{c}
e^{-i \arg{{\bf k}}}\\
1
\end{array}
\right)$, and ${u}_\tau^\prime({\bf k})=\bar{u}_\tau({\bf k})=\left[u_\tau({\bf k})\right]^*$. These states define Berry connections\cite{TKNN},  as ${\vec{\mathcal{A}}_\tau}={\vec{\bar{\mathcal{A}^\prime}}_\tau}=
-i \left[u_\tau({\bf k})\right]^{+}\vec\partial_k u_\tau({\bf k})$, ${\vec{\mathcal{A}^\prime}_\tau}={\vec{\bar{\mathcal{A}}}_\tau}=
-i \left[\bar{u}_\tau({\bf k})\right]^{+}\vec\partial_k \bar{u}_\tau({\bf k})$, and the Berry phases defined by contours $C_\tau$, $\tau=1,2,3$ encircling both, 
$P_\tau$ and  $P_\tau^\prime$ points, 
as $\gamma_\tau=\int_{C_\tau}d {\bf k} {\vec{\mathcal{A}}_\tau}$, and $\gamma^{\prime}_\tau=\int_{C_\tau}d {\bf k} {\vec{\mathcal{A}}^{\prime}_\tau}$.
The result of integration around each of these Dirac points yields
 $\gamma_\tau=\bar{\gamma}^\prime_\tau=-\bar{\gamma}_\tau=-{\gamma}^\prime_\tau=\pi$.

The corresponding Hamiltonian has the form

\bea
\label{S-Ham}
H=\frac{1}{2}\left(
\tiny{
\begin{array}{cccccc}
2 t \cos q_1&2 t( \cos q_2+i \sin q_3) &e^{i p_1} &e^{i p_2}-e^{i p_3} & e^{-i p_1}& e^{-i p_2}+e^{-i p_3}\\
2 t( \cos q_2-i \sin q_3) &-2 t \cos q_1& e^{i p_2}+e^{i p_3} & -e^{i p_1}& e^{-i p_2}-e^{-i p_3}&-e^{-i p_1} \\
e^{-i p_1}&e^{-i p_2}+e^{-i p_3} & -2 t \cos q_1&2 t(- \cos q_2+i \sin q_3) &e^{i p_1} &e^{i p_2}-e^{i p_3} \\
e^{-i p_2}-e^{-i p_3}&-e^{-i p_1}&-2 t( \cos q_2+i \sin q_3) & 2 J_2 \cos q_1& e^{i p_2}+e^{i p_3}&-e^{i p_1}  \\
 e^{i p_1}& e^{i p_2}-e^{i p_3} &e^{-i p_1} &e^{-i p_2}+e^{-i p_3} &-2 t \cos q_1& 2 t( \cos q_2-i \sin q_{3})\\
 e^{i p_2}+e^{i p_3}&-e^{i p_1} &e^{-i p_2}-e^{-i p_3} & -e^{-i p_1}&2 t( \cos q_2+i \sin q_3) & 2 t \cos q_1
\end{array}
}
\right).
\ena
At $t=J_2/J_1=0$, when we only have a small NN triangular lattice, the spectrum has a simple form. 
It becomes a spectrum of three Dirac pairs, having zeros at different points.
\bea
\label{J1}
E_{\tau,\bf k}=\pm \Big[3+\cos\big[2p_1+\frac{2\pi}{3}(\tau-2)\big]
+\cos\big[2p_2+\frac{2\pi}{3}(\tau-2)\big]-\cos\big[2p_3+\frac{2\pi}{3}(\tau-2)\big]\Big],
\;\; \tau=1,2,3 \nn
\ena 
The terms $2 \pi (\tau-2)/3$ in the arguments of the cosine functions appear due to the relative $2\pi/3$ rotations of three
NNN sublattices. 
The $\propto J_2$ terms in the Hamiltonian (\ref{S-Ham}) also have zeros at the same points, therefore
common Dirac points are $P_\tau$ and $\bar{P}_\tau$, $ \tau=1,2,3$. 
The chirality of point $\bar{P}_\tau\;\; \tau=1,2,3$, is opposite to that of
$P_\tau.$
The linear expansion of the spectrum around Dirac points gives anisotropic dispersion $
E^0_{\tau,\bf k}=\Big[\sum_{\mu}(1+ 3 t^2 )p^2_\mu -4 t p_\tau q_\tau \Big]^{1/2}.
$

\section{Self-consistency equations}

\subsection{Chern-Simons superconductor description of the $120^\circ$ phase}. 

Following the steps outlined in Ref.~\onlinecite{CSS}, we obtain the BdG Hamiltonian of the mean-field CS superconductor. 
In terms of the scalar order parameters $\Delta_{0,\bf k}$ and $\Delta_{3,\bf k}$, the self-consistency conditions take the closed form: 
\bea
\label{gap-eq}
\Delta_{0 \bf k}=\frac{2 \pi e}{3} \sum_{a=\pm,{\bf k' }}\sum_{\mu=1}^3\frac{\Delta_{3 \bf k'}}{{\bf k}^2 E^{(a)}_{\bf k'}} A^\mu_{\bf k-k'} k_\mu, \qquad
\Delta_{3 \bf k}=\pi e \sum_{a=\pm,{\bf k' }}\sum_{\mu=1}^3\frac{1}{E^{(a)}_{\bf k'}} 
\Big[u A^\mu_{\bf k-k'}k'_\mu
+ 2 w {\bf A}_{\bf k-k'}({\bf e}_3 q'_3+{\bf a}_3 p'_3)\Big],
\ena 
where $u=(a v_0-(1+t^2)\Delta_{0 \bf k'})$ and $w=t \Big(\frac{2 a (1+t^2)}{v_0}-\frac{4 t \Delta_{0 \bf k'}}{3}\Big)/\sqrt{3}$, with 
$v_0= \sqrt{(1+3t^2)^2-\frac{16}{3} t^2} $.
The  order parameters, Eq.~(\ref{gap-eq}), define the four-band  Bogolyubov spectra, $\pm E^{(a)}_{\tau, \bf k},\; a=\pm $, of gapped fermions as 
\bea
\label{spectr}
E^{(a)}_{\tau, \bf k}=\Big[\sum_{\mu=1}^3 p_\mu^2\big[(1+\Delta_{0 \bf k}^2)(1+3 t^2)- 
2 a v_0 \Delta_{0 \bf k}\big] 
- 4 t (1-\Delta_{0\bf k}^2) p_\tau q_\tau
+{\Delta^2_{3 \bf k}}\Big]^{1/2}.
\eea 
 At $t = 0$, the self-consistency equations are independent
of any continuous parameters (momentum cutoff is defined by the size of the Brillouin zone and is not a model
parameter). The existence of the solution within superconducting mean-field approach 
depends on the interaction strength, and thus on the CS charge, $e=1,3,5\ldots$. Here we have
one continuous parameter, $t$, and the anisotropy parameter
$g$ is set to zero.  At $t$ = 0,
the only solution that corresponds to e = 1 is the trivial
one, where there is no superconducting order. However, if
the CS-fermionization is realized with e = 3, a nontrivial
solution of gap equations emerges for $0\lesssim t\lesssim 0.089 $.

One can see that Eqs.~(\ref{gap-eq}) coincide with the analogous self-consistency relations of the CS superconductor on the honeycomb lattice
 at $t=0$, first derived in Ref.~\onlinecite{CSS}. This indicates the lattice independent `universal' character of CS superconductivity.

%%%%%%%%%%
\subsection{Chern-Simons superconductor description of the stripe phase}. 

Here we proceed with the fermionic description of the collinear stripe phase. 
The corresponding $\pi$-flux configuration is shown in Fig.~2d of the main text.
The BZ is still the same, but Dirac points now are
located at $Q=(\arccos(t),0)$ and $Q^{\prime}=(-\arccos(t),0)$. We see that the parity transformations, ${\cal P}_{x/y}$, transform one Dirac point to the other, indicating that the ground state does not support the degeneracy of the $120^\circ$ ordered state.  In the vicinity of these Dirac points 
the Hamiltonian acquires an especially simple form:
\bea
\label{ham-strip-2}
\!\!\!\!\!\!\!H_{\text{str}, 0}^{(Q)}({\bf k})&=& J_1\varepsilon \sum_{\bf k}\hat{f}^+_{\bf k,\alpha}\big[
v_y k_y \sigma^1_{\alpha\beta} -v_x k_x \sigma^3_{\alpha\beta}\big] \hat{f}_{\bf k,\beta},
\ena
with $v_x=\sqrt{1-t^2},\; v_y=\sqrt{\frac{3 (1-t)}{8}}(1-2 t)(1+t)$, and
$H_{\text{str},0}^{(Q')}({\bf k})=-H_{\text{str},0}^{(Q)}({\bf k})$.
The generated interaction vertex, $V^{\alpha\alpha',\beta\beta'}_{\bf q}$, in this case is similar to the one of the
$120\textdegree$ phase given below Eq.~(6) of the main text, but with 
$B^1_{\bf q}=-v_y A^x_{\bf q},\; B^2_{\bf q}=0,\; B^3_{\bf q}=-v_x A^y_{\bf q} $.
As in the case of the $120^\circ$ ordered state, here we also expect that $\Delta^{\alpha\alpha'}_{\bf k}$ has  a $p\pm ip$-wave symmetry, and the self-consistency relations are given by Eq.~(\ref{gap-eq}). 
The Bogolyubov mean-field treatment of the full Hamiltonian, $\mathcal{H}_{\text{str}}=H_{\text{str}, 0}^{(Q)}+H_{\text{str}, 0}^{(Q^\prime)}+H_{int}$, gives rise to a quasiparticle spectrum of the form 
\bea
\label{strip-spectr}
E^{(s,a)}_{\bf k}=\Big[(a v_x-v_y \Delta_{0 \bf k})^2k_x^2
+(a v_y-v_x \Delta_{0 \bf k})^2k_y^2
+{\Delta^2_{3 \bf k}}\Big]^{1/2}.
\eea

\section{Emergence of the Dirac Spin-Liquid}

As we see in Fig.~(5) of the main text, massless Dirac fermions emerge in the 
parameter interval  $0.089 \lesssim t\lesssim 0.116$.
Since the double degeneracy of the 
$120^\circ$ configuration of the the XX magnet implies a double degeneracy of the $\pi$-flux state of fermions on the triangular lattice, our approach indicates that the emergent Dirac spin-liquid state will also be doubly degenerate. 

The low-energy field theory in this regime 
appears to be  quite interesting. To understand its nature, one should integrate out fermionic degrees of freedom. 
In the interaction Hamiltonian, Eq.~(6) of the main text, only small values of momentum $q$ contribute to the formation of the order parameter $\Delta_{120^\circ}$. Momenta larger than 
a certain momentum cutoff,  $q \gtrsim \bar{Q}$, are irrelevant for the low-energy description of this ordered phase. 
When we approach criticality near $t\sim 0.089$, the fermion gap $\Delta_{120^\circ}$ vanishes, and critical fermions with large momenta yield the Maxwell term.
Indeed, the fermions do not fill any topological bands (e.g. we do not have a Chern insulator coupled to the gauge field) and integration over them will not result in generation of a Chern-Simons term~\cite{semenoff}. We rather have topologically trivial Dirac fermions coupled to the $U(1)$ "probe" field. Quantum fluctuations of fermions define an effective dynamics of the gauge field, 
which in the leading, one loop approximation yields a 2+1 dimensional Maxwell theory.
Thus at $0.116 \gtrsim t \gtrsim 0.089$,  one self-consistently obtains 
$N=6$ copies of Dirac fermions interacting with the induced $U(1)$ gauge field.

\end{widetext}
%%%%%%%%%%%%%%%
%%%%%%%%%%%%%%%


\begin{thebibliography}{99}

%%%%%%QSL
\bibitem{FA} P. Fazekas and P. W. Anderson, Phil. Mag. 30, 423 (1974).
\bibitem{balents}  L. Balents, Nature (London) 464, 199 (2010).
\bibitem{kitaev}  A.Y. Kitaev, Ann.Phys. 303, 2 (2003).
\bibitem{savary} L. Savary and L. Balents, Rep. Prog. Phys. 80, 016502 (2016).
\bibitem{norman} M. R. Norman, Rev. Mod. Phys. 88, 041002 (2016).
\bibitem{XGW} X. G. Wen, Rev. Mod. Phys. 89, 041004 (2017).
\bibitem{ng} Y. Zhou, K. Kanoda, and T.-K. Ng, Rev. Mod. Phys. 89, 025003 (2017).
\bibitem{KM} J. Knolle and R. Moessner, Annu. Rev. Condens. Matter Phys. 10, 451 (2019). 
\bibitem{senthil} C. Broholm, R. J. Cava, S. A. Kivelson, D. G. Nocera, M. R. Norman, T. Senthil,
"Quantum Spin Liquids," 	arXiv:1905.07040


%\bibitem{khveshch}
% D. V. Khveshchenko and P. B. Wiegmann, Mod. Phys. Lett. B 03, 1383 (1989);  {\em ibid.}, 04, 17 (1990).

%\bibitem{XGV} Xiao-Gang Wen, Phys. Rev. B 65, 165113 (2002).


%\bibitem{kivelson} H. Yao and S. A. Kivelson, Phys. Rev. Lett. 99, 247203 (2007). 



%\bibitem{berg} Y. Werman, S. Chatterjee, S. C. Morampudi, and E. Berg, Phys. Rev. X 8, 031064 (2018).

%\bibitem{shannon1} O. Benton, L.D.C. Jaubert, H. Yan, and Nic Shannon, 
%Nature Comm. 7, 11572 (2016).

%\bibitem{lu} Yuan-Ming Lu, Gil Young Cho, and Ashvin Vishwanath,
%Phys. Rev. B 96, 205150 (2017).
%\bibitem{pereira} G. Ferraz, F. B. Ramos, R. Egger, and R. G. Pereira,
%Phys. Rev. Lett. 123, 137202 (2019).


\bibitem{Andreson-1973} P. W. Anderson, Mater. Res. Bull. 8, 153 (1973).

\bibitem{ss} S. Sachdev, Phys. Rev. B 45, 12377 (1992).
\bibitem{Moessner1} R. Moessner and S. L. Sondhi, Phys. Rev. Lett. 86, 1881 (2001).
\bibitem{moess} R. Moessner and S. L. Sondhi, Prog. Theor. Phys. Supplement 145, 37 (2002).
\bibitem{AV} F. Wang and A. Vishwanath, Phys. Rev. B 74, 174423 (2006).
\bibitem{wietek} A. Wietek and A. M. L{\"a}uchli, Phys. Rev. B 95, 035141 (2017).




%%%%%Theory with SO coupling

\bibitem{chen17} Yao-Dong Li, Yuan-Ming Lu, and Gang Chen,
Phys. Rev. B 96, 054445 (2017).

\bibitem{chen18} Yao-Dong Li, Yao Shen, Yuesheng Li, Jun Zhao, and Gang Chen,
Phys. Rev. B 97 125105 (2018).

\bibitem{starykh} Z.-X. Luo, E. Lake, J.-W. Mei, and O. A. Starykh,
Phys. Rev. Lett. {\bf 120}, 037204 (2018).

%%%%%%%%%%%%%EXPERIMENTS on YbMgGaO4

\bibitem{shen} Y. Shen, Y.-D. Li, H. Wo, Y. Li, S. Shen, B. Pan, Q. Wang, H. C.
Walker, P. Steffens, M. Boehm, Y. Hao, D. L. Quintero-Castro, L. W. Harriger, M. D. Frontzek, L. Hao, S. Meng, Q. Zhang, G. Chen, and J. Zhao, Nature (London) 540, 559 (2016).

\bibitem{yli} Yuesheng Li, Devashibhai Adroja, Pabitra K. Biswas, Peter J. Baker, Qian Zhang, Juanjuan Liu, Alexander A. Tsirlin, Philipp Gegenwart, and Qingming Zhang, Phys. Rev. Lett. 117, 097201 (2016).

\bibitem{paddison} J. A. M. Paddison, M. Daum, Z. L. Dun, G. Ehlers, Y. Liu,
M. B. Stone, H. D. Zhou, and M. Mourigal, Nat. Phys. (2017),
doi:10.1038/nphys3971.



%%%%%% numerics 

%%%%VMC
\bibitem{vmc1} R. Kaneko, S. Morita, and M. Imada, J. Phys. Soc. Jpn. 83,
093707 (2014).

\bibitem{becca} Y. Iqbal, W. Hu, R. Thomale, D. Poilblanc, and F. Becca, Phys. Rev. B {\bf 93}, 144411 (2016).
\bibitem{Balents-2018}  J. Iaconis, C. Liu, G. B. Halasz, and L. Balents, SciPost Phys. {\bf 4}, 003 (2018).

\bibitem{becca-prx} Francesco Ferrari and Federico Becca,
Phys. Rev. X 9, 031026 (2019).

%%%%%%%%%%%%%%%%%%

%%%%%DMRG

\bibitem{White-2015} Z. Zhu and S. R. White, Phys. Rev. B 92, 041105 (2015).

\bibitem{Sheng-2015} W.-J. Hu, S.-S. Gong, W. Zhu, and D. N. Sheng, Phys. Rev. B
92, 140403(R) (2015).
\bibitem{crit} Z. Zhu, P. A. Maksimov, S. R. White, and A. L. Chernyshev,
Phys. Rev. Lett. 119, 157201 (2017).

\bibitem{xxz} Z. Zhu, P. A. Maksimov, S. R. White, and A. L. Chernyshev,
Phys. Rev. Lett. 120, 207203 (2018).


%%%%% WE

\bibitem{CSS} T. A. Sedrakyan,  V. M. Galitski, and A. Kamenev, Phys. Rev. B {\bf 95}, 094511 (2017).

%%%%%%%%%%%%Classical

\bibitem{Lee} D. H. Lee, J. D. Joannopoulos, J. W. Negele, and D. P. Landau
Phys. Rev. Lett. {\bf 52}, 433 (1984). 

\bibitem{Shiba} S. Miyashita and H.~Shiba,  J. Phys. Soc. Jap. {\bf 53},  1145 (1984).

\bibitem{korshunov} S. E. Korshunov, Physics - Uspekhi {\bf 49} (3) 225 - 262 (2006).






%%%%%%%%%%%%

\bibitem{Chubukov-17} M. Ye and A. V. Chubukov, Phys. Rev. B 95, 014425 (2017).
\bibitem{Chubukov-18} M. Ye and A. V. Chubukov, Phys. Rev. B 97, 245112 (2018).

%%%%%%%%%%%%

%\bibitem{nakamura}  Tota Nakamura and Shin-ichi Endoh, J. Phys. Soc. Jpn. 71, 2113 (2002).

%\bibitem{danu} B. Danu, G. Nambiar, and R. Ganesh, Phys. Rev. B 94, 094438 (2016).
%%%%%%%%%%%%

\bibitem{WWS} R. Wang, B. Wang, and T. A. Sedrakyan, Phys. Rev. B {\bf 98}, 064402 (2018).

\bibitem{SKG2} T. A. Sedrakyan,  L. I. Glazman, and A. Kamenev, Phys. Rev. Lett. {\bf 114}, 037203 (2015).

\bibitem{SKG} T. A. Sedrakyan,  A. Kamenev, and L. I. Glazman, Phys. Rev. A {\bf 86}, 063639 (2012).

\bibitem{SGK1} T. A. Sedrakyan,  L. I. Glazman, and A. Kamenev,  Phys. Rev. B {\bf 89}, 201112(R) (2014).

\bibitem{maiti} S. Maiti and T. Sedrakyan, Phys. Rev. B {\bf 99}, 174418 (2019).


%\bibitem{supplement} For details see Appendix.


\bibitem{villain} J. Villain, J. Phys. France 38, 4 (1977) 385-39.

\bibitem{KW}  H. Kawamura, Phys. Rev. B 38, 4916 (1988).



%%%%%%%SKG2 SKG SGK1





\bibitem{hermele} Michael Hermele, T. Senthil, Matthew P. A. Fisher, Patrick A. Lee, Naoto Nagaosa, and Xiao-Gang Wen,
Phys. Rev. B 70, 214437 (2004).


\bibitem{XGV} Xiao-Gang Wen, Phys. Rev. B 65, 165113 (2002).

\bibitem{Bieri} S. Bieri, C. Lhuillier, and L. Messio,
Phys. Rev. B 93, 094437 (2016).

\bibitem{Maleyev} S.V. Maleyev, Physica B 345, 119-123 (2004).

\bibitem{Simonet} V. Simonet, M. Loire, and R. Ballou, Eur. Phys. J. Special Topics 213, 5-36 (2012).

\bibitem{ER} E. Ressouche, "Polarized neutron diffraction," Collection SFN 13, 02002 (2014).


%\bibitem{SKG2} T. A. Sedrakyan,  L. I. Glazman, and A. Kamenev, Phys. Rev. Lett. {\bf 114}, 037203 (2015).

%\bibitem{CSS} T. A. Sedrakyan,  V. M. Galitski, and A. Kamenev, Phys. Rev. B {\bf 95}, 094511 (2017).

\bibitem{TKNN}
D. J. Thouless, M. Kohmoto, M. P. Nightingale, and M. P. M. den Nijs, Phys. Rev. Lett., {\bf 49},
405 (1982).

\bibitem{semenoff}  G. W. Semenoff, Phys. Rev. Lett. {\bf 53}, 2449 (1984).

\end{thebibliography}
\end{document}